\documentclass[a4paper,showpacs]{revtex4}
\usepackage{amstext,amsfonts,amsmath,amssymb,amsthm}
\usepackage{color}
\usepackage{epsf}
\usepackage{graphics}

\newfont{\gl}{eufm10 scaled \magstep1} 

\def\simlim{\mathop{\sim}}

\newcommand{\beq}{\begin{equation}}
\newcommand{\eeq}{\end{equation}}

\newcommand{\beqs}{\begin{equation*}}
\newcommand{\eeqs}{\end{equation*}}

\begin{document}

\title{Neutrino mass and Extreme Value Distributions
 in $\beta$-decay}
 \author{J. G. Esteve\footnote{Corresponding author.}}
 \email{esteve@unizar.es}
 \affiliation{Departamento de F\'{\i}sica Te\'orica, Universidad de Zaragoza,
50009 Zaragoza, Spain}
\affiliation{Instituto de Biocomputaci\'on y F\'{\i}sica de Sistemas
Complejos (BIFI), 50009 Zaragoza, Spain}
  \author{Fernando Falceto }
\email{ falceto@unizar.es}
 \affiliation{Departamento de F\'{\i}sica Te\'orica, Universidad de Zaragoza,
50009 Zaragoza, Spain}
\affiliation{Instituto de Biocomputaci\'on y F\'{\i}sica de Sistemas
Complejos (BIFI), 50009 Zaragoza, Spain}

\date{\today}

\begin{abstract}

\noindent
We propose to apply the Extreme Value Theory for distributions 
with compact support to the estimation of the neutrino mass 
from the energy spectrum of electrons in $\beta$-decay.
Using these techniques the dependence of the results on the mass
of the neutrino is considerably enhanced increasing the sensitivity
of the experiment. We discuss how these tools could be applied to the
present proposals like the KATRIN experiment.
\end{abstract}


\maketitle

\section{Introduction}
Neutrino masses play an important role in different areas of Physics ranging
from Nuclear to High Energy Physics or Astrophysics. Up to date, however,
we only know upper bounds to their values and their determination remains 
an outstanding problem in Physics.
One of the most promising procedures to unravel the effective mass of 
the electronic (anti)neutrino is to study the electronic energy spectrum in the
$\beta$-decay, where the maximum energy allowed to the electron depends on
the mass of the (anti)neutrinos (in what follows we will not differentiate between
the masses of neutrino and antineutrino, as we  assume an exact 
CPT symmetry). However,
the ratio of the effective electron-neutrino mass $m_\nu$ to the electron mass $m_e$ is 
very small, $m_\nu / m_e < 10^{-5}$ (\cite{PDG} -\cite{Troisk}), 
and the probability of emission of the electron
depends at least quadratically on this quotient.
This implies that the spectrum of the emitted electrons is 
very little affected by the value of $m_\nu$.
Only near the end point of the spectrum, where the energy
of the emitted antineutrino is close to $m_\nu c^2$, there is a stronger
dependence in the neutrino mass as the highest energy
of the electron varies linearly with $m_\nu c^2$. 
This adds new difficulties to measure the neutrino mass in $\beta$-decay, 
since one must explore the region very near the end point, 
where the occurrence of an event has a very small probability.

For these reasons, it is important to find new strategies 
to analyse the data with higher sensitivity in this part of 
the electron spectrum. This goal can be
achieved with the help of the Extreme Value Theory (EVT) for random variables 
of  compact support \cite{estfal}, as it is described below.

EVT studies the probability distribution of the maximum of $n$ independent, 
identically distributed random variables (\cite{HannFerreira} - \cite{mohrav}).
Its interest in connection to the 
problem of determining the neutrino mass is clear if one considers that,
as mentioned before, the influence of the neutrino mass on the energy distribution 
of the electronic emission, is maximal at the upper bound. Then, if we 
focus on the electrons emitted with the highest energy among $n$ emissions
we should expect a deeper insight of the region near the upper bound 
of the spectrum. In the next sections we will 
introduce the main tools in EVT
and we will present quantitative results for the Tritium 
$\beta$-decay showing the enhancement in the sensitivity of the experiment.

\section{Extreme statistics}
Let us  consider a normalised probability density $\rho(x)$ 
with support in $[0,1]$, i.e. 
$\rho(x)=0$ if $x\not\in[0,1]$,  and
$\mu(x)$ its cumulative distribution function:
$$\mu(x)=\int_0^x \rho(s) ds.$$
The distribution function of the maximum of $n$ independent
identically distributed (i.i.d.) 
random variables with probability density $\rho(x)$ is then:
\begin{equation}
 \mu_n(x)=\mu(x)^n,
\end{equation}
as it can be easily argued. In fact, note that $\mu(x)$ represents
the probability of the random variable taking a value smaller than
$x$. Therefore, the probability of $n$ i.i.d. random variables 
taking all of them a value smaller than $x$ is the $n$th power of the former \cite{noiid}. 
Hence, the probability density is:
\begin{equation}
 \rho_n(x)=\frac{d}{dx} \mu_n(x)=n \mu(x)^{n-1} \rho(x).
\end{equation}

We are interested in the large $n$ limit where 
it is evident that $\rho_n(x)$ squeezes 
at the end point $x=1$ independently of the initial probability density
(provided $x=1$ is in the support of $\rho$).
Therefore, we lose any sign of the initial distribution.

A way to overcome this situation is to renormalise, at every step,
the random variable. As it is shown in \cite{estfal}
a convenient renormalisation is to transform 
the variable $x$ into $x'=x^{n^{-\beta}}$. 
This transformation has the virtue of
preserving the support $[0,1]$, which means that the new random variable $x'$
takes the same range of values as the original one.  Also, if we chose the 
appropriate $\beta$, the limit of infinite $n$ provides a non trivial
probability distribution.

The problem of choosing the appropriate value for $\beta$ was addressed 
in \cite{estfal} using Renormalisation Group techniques. There
it is shown that $\beta$ depends solely on the behaviour of $\rho(x)$ in the 
vicinity of $x=1$ 
More specifically, assuming that 
\begin{eqnarray}\label{behaviour}
\rho(x)\simlim_{x\to1}& \gamma(1-x)^{\alpha-1},\nonumber\\
\end{eqnarray}
(where  $\alpha>0$ in order to have a normalisable density) the appropriate 
transformation is implemented with $\beta=1/\alpha$.

Once the right transformation has been chosen 
one can safely take the large $n$ limit of  $\mu_n(x^{n^{-1/\alpha}})$, 
namely:
\begin{equation}
 M(x)=\lim_{n\to\infty} \mu_n(x^{n^{-1/\alpha}})= e^{-\lambda(-\log(x))^\alpha},
 \label{limdis}
\end{equation}
where $\lambda=\gamma/\alpha$ and, as shown in \cite{Falceto},
the limit of probabilities distributions 
should be understood in the weak sense (or in law).
To derive the result in \ref{limdis} it is enough to realise that
the large $n$ limit corresponds to a fixed point of the Renormalisation Group 
transformation, i.e it should satisfy
\begin{equation}
M(x^{n^{-1/\alpha}})^n=M(x).
 \label{fixedpoint}
\end{equation}
In terms of the function $\Phi(y)=\log M(e^{-y})$ the previous equation reads
$$n \Phi(n^{-1/\alpha} y)=\Phi(y),$$
which implies that $\Phi$ is homogeneous of degree $\alpha$ and therefore
\ref{limdis} follows (the coefficient $\lambda$ can be obtained from the 
behaviour of $\mu(x)$ at $x=1$).

Observe that the limiting probability distribution
depends only on the two parameters $\alpha$ and $\gamma$ 
that determine the behaviour of $\rho$ at the maximum of its 
support. All other details of the initial distribution are 
swept out. Note also that the limit in
(\ref{limdis}) is related to the  Fr\'echet distribution by changing to 
the variable $y=-\log x$. The latter distribution is one of the possible limits in the Gnedenko's extreme value theorem  \cite{Gnedenko}.

If we had considered a different rescaling 
($\beta\not=1/\alpha$) then
the  probability density, in the limit of infinite $n$, 
would have been trivial in the sense that it accumulates either
at $0$ or at $1$. In fact, in this case we have
\begin{equation}
\lim_{n\to\infty} \mu(x^{n^{-\beta}})^n =
\begin{cases}
H(x)& \mbox{if }\beta>1/\alpha\cr
H(x-1)& \mbox{if }\beta<1/\alpha
\end{cases}
\end{equation}
with $H$ the Heaviside step function. 

It should be stressed that these completely opposite limits 
hold only if $n$ goes strictly to $\infty$ which, of course, 
is not attainable experimentally. However, one expects that
when $n$ is large enough (in the sense that we will precise in
the next section) the renormalised distribution depends
critically on its behaviour at the maximum of its support.

These properties are key to our purposes, 
due to the fact that the energy spectrum
of the electron emitted in the $\beta$ decay changes drastically
at the upper limit depending whether the neutrino is massless
or not. Namely, if the neutrino were massless the
probability density for the electronic energy would behave
like $(E_{\rm  max}-E)^2$ ($\alpha=3$).
On the other hand, if it were massive the behaviour would be  
$(E_{\rm max}-E)^{1/2}$ ($\alpha=3/2$). 
Therefore, we expect that the sensitivity of the experiment
to the neutrino mass should be very much enhanced when
the results are analysed with the tools provided by EVT. 

\section{Application to Tritium $\beta$-decay} 
In this section we shall apply the previous ideas to the specific problem of 
determining the neutrino mass in $\beta$-decay experiments.
Although the method is general, we will focus 
on the $\beta$-decay of Tritium.
This process is specially suited for the purpose of 
determining the neutrino mass
for two reasons. 
First, it is a super allowed decay ($12.3$ years of half-live) and, 
consequently, the nuclear matrix element
does not depend on the energy of the emitted electron.
Second, it has a relatively 
low end point energy, which increases the sensitivity of the experiment
to the neutrino mass 
(see \cite{Otten} for a review on the neutrino mass limit from the Tritium beta decay).

The probability of emission of an electron,
with total energy $E$ ranging from 
$m_e c^2$ to $E_0-m_\nu c^2$, is given by
\begin{equation}
p_{_{\rm e}}(E)=C_{_{\rm e}}G(E)\, (E_0-E)\,[(E_0-E)^2-m_\nu^2c^2]^{1/2},
\label{rho}
\end{equation}
where $C_{_ {\rm e}}$ is a normalisation constant, $E_0$ the total available energy, i. e. 
$E_0=\epsilon_0 + m_e c^2$ with $\epsilon_0$ the end point energy,
and 
$$G(E)=\frac{E^2}{1-{\rm e}^{-\frac{2\pi}{137}\frac{E}{\sqrt{E^2-m_{\rm e}^2c^4}}}},$$
is related to the Fermi function which accounts for the interaction with the nuclear Coulomb field. 
In the case of atomic Tritium we shall take
$\epsilon_0=18.560$ KeV \cite{Otten}.

The constant $m_\nu^2$ that appears in \ref{rho}
is, actually, the squared effective mass of 
the electron antineutrino, which should be expressed
in terms of the weighted average of the mass eigenstates $m_i$ ($i=1,2,3$) as 
$m_\nu^2=\sum_{i=1,3} |U_{ei}|^2 m_i^2$ being $|U_{ei}|^2$ 
the weights known from the neutrino oscillation experiments.
For the sake of clarity we have neglected in (\ref{rho})
the sum over final states. This does not affect
the conclusions and it can be implemented numerically.

It will be convenient to introduce dimensionless, rescaled variables for the
electron kinetic energy and the neutrino mass. Namely, define 
\begin{equation}
x=\frac{E-m_{\rm e}c^2}{E_0-m_{\rm e}c^2},\quad y=\frac{m_\nu c^2}{E_0-m_{\rm e}c^2},
\label{cambio}
\end{equation}
 and the probability density in terms of $x$ is
\begin{equation}\label{density}
\rho(x)=C\,F(x)\, (1-x)[(1-x)^2-y^2]^{1/2}
\end{equation}
where $0<x<1-y$, $C$ is the new normalisation constant and 
$F(x)= G(\,x(E_0-m_{\rm e}c^2)+m_{\rm e}c^2\,)$. 
Now by monitoring the energy of the electrons in the $\beta$-decay of Tritium
we could estimate the probability density $\rho$ and deduce 
the value of $y$. The problem is that, as we mentioned before, 
the dominant correction to the limit of zero neutrino mass goes like $y^2$ 
\cite{fn}
and, given the small upper bound value for $y\lesssim 10^{-5}$, the corrections due to
the mass of the neutrino are very small. As we shall see, the EVT
for $n$ events provides an enhancement of the correction by a factor $n^{2/3}$
making it easier to detect.

The experimental set up should focus on the electron of highest energy among $n$ 
decays. That is, given a set of data corresponding to $n$ decays, we keep only
those corresponding to the emitted electron with the maximum energy; 
we call this event an extreme value decay. 
Repeating this selection process $N$ times over different sets of $n$ decays,
we can obtain the distribution function for the extreme value decays
$\mu_n^{exp} (x)$ and from it $M^ {exp}(x)=\mu_n^{exp} (x^{n^{1/3}})$ which can be
fitted to the rescaled limit distribution (\ref{limdis}) obtained for a given value of $m_\nu$ 
(or equivalently with $P(x)=\frac{d}{dx} M(x)$ ).
 Note that the total number of
disintegrations needed is $n N$.
The number $n$ of decays that we consider at each step can be determined 
if  we know how many atoms of Tritium we have in the sample and the time 
spent recording the energy of the emitted electrons. 
Moreover, we do not have to register all the $n$ events, because as we are finally 
interested in the highest energy emitted electron, a very stringent cut-off 
can considerably reduce the number of decays that one has to analyse. 

At present, 
the KATRIN experiment \cite{KATRIN1,KATRIN2}
is planned to operate with a $10^{11}$ Bq. source and a running time 
of $10$ years that gives a total number of $3\,\times 10^{19}$ decays, 
then a reasonable choice for an EVT analysis would be 
$N\in [10^ 6, 10^ 8]$ and
$n\in [10^ {11}, 10^ {13}]$. In what 
follows we shall use $n=10^ {12}$ as a realistic value, this implies that we 
can record an extreme event every $10$ seconds.

The cumulative distribution for the highest energy events is
$\mu_n(x)=\mu(x)^n$, where $\mu$ is the cumulative
distribution function for the probability density in (\ref{density}).
In order to get a non trivial limit (in the case of zero neutrino mass)
we shall rescale the variable and define $M_n(x)=\mu_n(x^{n^{-1/3}})$.
To estimate the size of the corrections due to the non zero neutrino mass
we first expand $\rho$ in powers of $y$,
$$\rho(x)= a(x) (1-x)^2+y^2 b(x)+ O(y^3)$$
where 
\begin{eqnarray*}
&&a(x)=Q^{-1} F(x),\cr
&&b(x)=[(1-x)^2Q^{-1}R-1]\frac{Q^{-1}F(x)}2,
\end{eqnarray*}
with
$Q=\int_0^1 F(x)\, (1-x)^2 {\rm d}x$ and  $R=\int_0^1F(x)\, {\rm d}x.$
Then the cumulative distribution $\mu$ can be similarly written
$$\mu(x)= 1-A(x)\, (1-x)^3-y^2 B(x)\, (1-x)+ O(y^3).$$
For the moment we do not specify the concrete form of $A(x)$ and $B(x)$.

Now we can easily expand the logarithm of $M_n(x)=\mu(x^{n^{-1/3}})^n$:
$$\log M_n(x)=A(x^{n^{-1/3}}) \log^3 x + n^{2/3} y^2 B(x^{n^{-1/3}})\log x+\dots$$
where the dots represent
subdominant contributions for large $n$ and small $y$
and we have used $1-x^{n^{-1/3}}=-n^{-1/3}\log x+\dots$.
Therefore, taking the large $n$ limit (with $n^{1/3} y< 1$) in the arguments of $A$ and $B$
and exponentiating the previous expression we obtain
\begin{equation}\label{expansion}
M_n(x)={\rm e}^{\lambda \log^3 x}(1-\frac32n^{2/3}y^2 \lambda\log x+\dots).
\end{equation}
Where $\lambda=A(1)=(1/6) \mu'''(0)|_{y=0}$ and, given the relation
between $\mu$ and $\rho$, we have
$\lambda=(1/6) \rho''(0)\vert_{y=0}=a(1)/3\approx 2.04208$.

The expansion for the probability density is obtained by taking the 
derivative of (\ref{expansion}) and it reads
\begin{eqnarray}\label{expdensity}
P_n(x)&=&P(x)\Big(1-
\frac12n^{2/3}y^2 \big(3\lambda\log x+\log^{-2}x\big)+\dots\Big),\nonumber
\end{eqnarray}
where $P(x)$ is the probability density in the limit of zero neutrino mass, i.e.
$$P(x)=3\lambda\frac{\log^2x}x{\rm e}^{\lambda \log^3 x}.$$

The  expressions above are valid provided $n\gg1$ with $n^{1/3} y<1$ which is the real situation, 
 given the actual upper bounds for $m_\nu $ and the accessible experimental 
 values for $n$ (of the order of $10^{12}$, as we have seen before)(\cite{KATRIN1}-\cite{KATRIN2}).
If we had considered the large $n$ limit
in a more strict sense, $n\gg1$ and $n^{1/3} y \gg1$, 
the results would have changed and the difference between the massless and 
massive case would have been more dramatic. In fact, after performing the 
appropriate rescaling, we would have obtained for {\it very} large $n$:
\begin{eqnarray}
 \log M_n(x)\propto 
\begin{cases}
-|\log x|^3&\quad (m_\nu=0)\\
-|\log x|^{3/2}&\quad (m_\nu\ne 0). 
\end{cases}
\label{nbig}
\end{eqnarray}
However, as we argued before, in the present situation this limit is not attainable.
Therefore, we shall restrict ourselves to the case when $n^{1/3} y <1$
using the expression (\ref{expansion}) for $M_n(x)$.

 \begin{figure}[h!]
\epsfxsize=7cm
\begin{center}
\leavevmode   
\epsffile{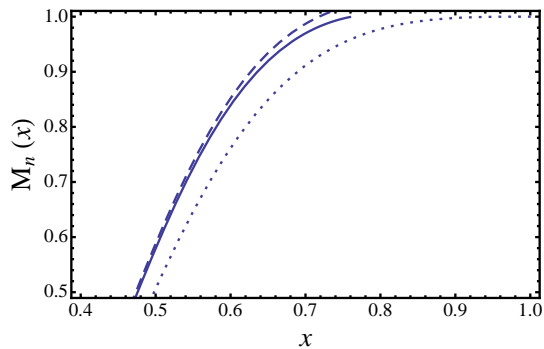}
\end{center}
\caption{Comparison between the exact value of $M_n(x)$ (continuous line) and the
perturbative value, up to second order in $y$, obtained from (\ref{expansion})
for $n=10^{12}$ and $m_\nu/m_e= 10^{-6}$. The dotted line represents $M_n(x)$ for a massless neutrino.}
\label{fig1}
\end{figure}

In our analysis of the Tritium $\beta$-decay we address 
the pos\-si\-bil\-i\-ty of measuring a neutrino mass of the order 
$m_\nu/m_e=10^{-6}$. In Fig. \ref{fig1} we show a comparison 
between the exact values of $M_n(x)$ for $y=0$ 
and for  $y= 2.8\times 10^{-5}$ ($m_\nu=10^{-6} m_e$). 
We also show the curve obtained using the perturbative 
expansion (\ref{expansion}) with $n^{2/3} y^2=0.078$. 
We can see there  that the perturbative expansion fits 
well with the exact value.
  
 \begin{figure}[h!]
\epsfxsize=7cm
\begin{center}
\leavevmode   
\epsffile{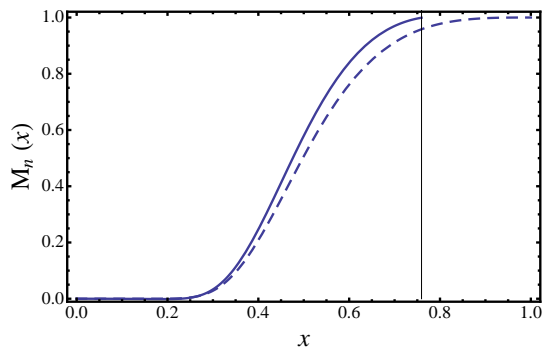}
\end{center}
\caption{ $M_n(x)$ for $n=10^{12}$ and  $m_\nu/m_e= 10^{-6}$ (continuous line),
and for $m_\nu=0$ (discontinuous line). The  vertical line indicates the value of
$(1-y)^{n^{-1/3}}$}
\label{fig2}
\end{figure}
 
Next, in Fig. \ref{fig2}  we plot the cumulative distribution $M_n(x)$
 for $n=10^{12}$, $m_\nu=0$ and $m_\nu c^2=0.5$ eV. Note that for this value of
 $n$, the distributions
  corresponding to $m_\nu c^2 = 0.5$ eV and $m_\nu=0$ are clearly separated.
\begin{figure}[h!]
\epsfxsize=7cm
\begin{center}
\leavevmode   
\epsffile{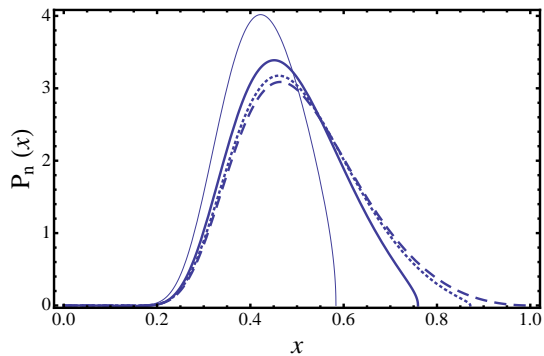}
\end{center}
\caption{$P_n(x)$ for $n=10^{12}$ and  $m_\nu c^2= 1$ eV (thinnest line),
 $m_\nu c^2= 0.5 $ eV (continuous thicker line), $m_\nu c^2= 0.25 $ eV (dotted line)
and $m_\nu=0$ (discontinuous line). }
\label{fig3}
\end{figure}

In Fig. \ref{fig3} we compare the probability distribution $P_n(x)$ for $n=10^{12}$ and 
different values of $m_\nu$. 
The graph illustrates how the bound 
on the neutrino mass, 
for this $n$, 
can be lowered by 
one order of magnitude with respect to its present value. 
However, it would be very difficult to go one order of magnitude further, as 
it would imply to increase $n$ by a factor $10^3$.
 
We end our discussion by studying the expectation value of the variable 
$x^{n^{1/3}}$, that is also sensitive to the neutrino mass. In this case we 
also discuss how the precision in the measurement of the electronic energy
affects the determination of the neutrino mass. We have:
 \begin{eqnarray*}
 \langle x^{n^{1/3}}\rangle&=&\int_0^{1-y} {x^{n^{1/3}} \frac{d}{dx} \mu_n(x) dx} \\
  &=& \langle x^{n^{1/3}}\rangle_{y=0} - \frac32n^{2/3} y^2 \lambda \int_0^\infty {\rm e}^{-\lambda t^3-t}t dt+\dots, 
 \end{eqnarray*}
where $\langle x^{n^{1/3}}\rangle_{y=0}$ is the mean value for $m_\nu=0$. In our case
\begin{equation}\label{expectation}
 \langle x^{n^{1/3}}\rangle \approx 0.510864 -0.496771\, n^{2/3} y^2 +\dots,
\end{equation}
and the dots mean corrections of order $n^{-1/3}$ and $n^{4/3} y^4$.
 
Assume now an indetermination $\Delta E$ in the measurement of the 
electronic energy,
this is related to the indetermination of the neutrino mass 
($y$ in the adimensional variables) by the following expression,
$$\Delta E=0.993542\times(E_0-m_{e} c^2)\frac{n^{1/3}y^2}{\langle x^{n^{1/3}}\rangle}\frac{\Delta y}y$$
which one easily gets from (\ref{expectation}),  
Therefore, assuming $m_\nu\sim 0.5$ eV, we can determine it, in an individual measurement, 
within $50\%$ of relative precision provided the energy resolution is 
$\Delta E\sim 0.88$eV
and $n=2\times 10^{13}$  
(an account every 200 seconds in an experiment with $10^{11}$ Bq).
This estimates are close to the values that are expected from future 
experiments \cite{KATRIN1,KATRIN2}.

\section{conclusions}
In this paper, we briefly presented the main tools of 
Extreme Value Statistics for distributions of compact support.
A subject that can be useful for analysing the electron spectrum 
in the $\beta$-decay of tritium, in order to determine an upper
limit for the electronic neutrino. Actually, the main difference in the 
spectrum with a massive or a massless neutrino is its behaviour
at the tail of the distribution, a region that can be deeply explored
thanks to the extreme value theory.

We have performed an analytic study of the expectation variable of the 
renormalised energy of the fastest emitted electron among $n$ 
disintegrations. We have shown that the expectation variable
depends linearly on $n^{2/3}y^2$ and we have determined the 
coefficients. In this way we show that the dependence
on the square mass of the neutrino (in dimensionless units) $y^2$
is enhanced by the factor $n^{2/3}$, which increases the sensitivity
of the experiment. For instance, in the present experimental facilities,
we can reach $n\approx 10^{12}$, which means that neutrino mass 
upper bounds of the order of $0.25$ eV  could be obtained.
In conclusion, we believe that extreme value theory can be useful
to analyse the electronic energy spectrum in $\beta$-decay and it is worth 
implementing in present and future experiments. 

\vskip 1cm

\noindent{\bf Acknowledgements:}  Research partially supported by
grants 2012-E24/2, DGIID-DGA and FPA2012-35453, Ministerio de Industria y
Competitividad (Spain). We thank Amalio Fern\'andez-Pacheco for comments.



\begin{thebibliography}{99}
\bibitem{fn}{Note that, although apparently $\rho$ depends quadratically on $y$, 
there could be a linear contribution coming from the normalisation constant. 
However, one can deduce that this contribution  is absent due to the fact 
that $\rho$ vanishes in the upper limit $x=1-y$.}
\end{thebibliography}
\end{document}